\documentstyle{elsart}

\begin{document}
\begin{frontmatter}

\title{Topological Dynamics and Grand Unified Theory}

\author[jhu]{Marco Spaans\thanksref{em1}}

\address[jhu]{Physics \& Astronomy Department, Johns Hopkins University,
3400 North Charles Street, Baltimore, MD 21218-2686, USA}
Email: spaans@pha.jhu.edu

\thanks[em1]{spaans@pha.jhu.edu}

\begin{abstract}

A formalism is presented to construct a non-perturbative Grand
Unified Theory when gravitational Planck-scale phenomena are included.
The fundamental object on the Planck scale is the three-torus $T^3$
from which the known properties of superstrings, such as the
geometric action and duality, follow directly. The low energy theory is
11-dimensional and compactification to a Lorentzian four-manifold is
an automatic feature of the unified model. In particular, the
simply-connected K3 Calabi-Yau manifold follows naturally from the
model and provides a direct link with M-theory. The high energy
theory is formulated on a $T^3$ lattice with handles which exhibits the
necessary symmetry groups for the standard model, and yields a consistent
amplitude for the cosmic microwave background fluctuations.
The equation of motion for the supersymmetric, unified theory is derived and
leads to the Higgs field. This formulation predicts a remnant, scalar
topological defect of mass $\le 9m_{\rm Planck}/46$ from the Planck epoch,
which is a candidate for dark matter.\hfill\break
Finally, it is shown that if
the universe is quantum-mechanical, its spatial dimension is equal to
three and the laws of nature are Lorentz invariant when gravity can be
neglected.
\hfill\break\noindent
Keywords: general relativity --- quantum cosmology\hfill\break

\end{abstract}
\end{frontmatter}

\section{Introduction}

One of the outstanding questions in quantum cosmology and particle physics
is the unification of gravity with the electro-weak and strong interactions.
Much effort has been devoted in the past years to formulate a purely
geometrical and topological theory for both types of interactions[1,2].
This work aims at extending the topological dynamics approach presented in [3]
(paperI from here on) to include non-gravitational interactions and to
construct a Quantum-gravitational Grand Unified Theory (QGUT).

The main results of paperI are as follows:\hfill\break
The properties of space-time topology are governed by homotopically
inequivalent loops in the {\it prime} manifolds $S^3$,
$S^1\times S^2$, and $T^3$. This is the only set which assures Lorentz
invariance and the superposition principle for all times, and thus provides a
natural boundary condition for the universe. The dynamics of the theory are
determined by the loop creation ($T^\dagger$) and loop annihilation ($T$)
operators. On the Planck scale the quantum
foam has the unique structure of a lattice of three-tori with 4 homotopically
inequivalent paths joining where the $T^3$ are connected through
three-ball surgery. The existence of this four-fold symmetry is called
{\it multiplication} and leads to O(n) and SU(n) gauge groups as well as
black hole entropy. The number of degrees of freedom, i.e.~the prime quanta,
associated with the three-sphere,
the handle manifold, and the three-torus are 1, 3, and 7, respectively.
During the Planck epoch, $S^1\times S^2$ handles can attach themselves to
the $T^3$ lattice and lead to supersymmetric interactions.

In a way one can say that the quantum-mechanical formulation of the Equivalence
Principle is found in the occurrence of multiplication, which accounts for the
homotopic Planck scale degrees of freedom not treated in general relativity.
Also note that Wheeler's
argument for a ``defective gene'' in $T^3$ related to the initial value problem
of general relativity[4] is avoided, because metrics and connections
are defined
on the intersection points of the lattice. Finally, it was found that the
cosmological constant is very small and proportional to the number of
macroscopic black holes at the current epoch.
This result follows from the generalization of Mach's principle in which
local topology change is a consequence of global changes in the matter degrees
of freedom and vice versa. Such a result
is consistent with 't Hooft's ``Naturalness Principle'' and the fact that the
symmetry of the Einstein equations is not increased when $\Lambda = 0$.
Since life forms can only exist in a universe where the bulk of matter is
not in the form of black holes, it follows from the Anthropic Principle that
we observe a small cosmological constant.
A further consequence is that massive isolated black holes still grow due to
quantum fluctuations. This steady increase in entropy, independent of the
matter degrees of freedom, yields a natural arrow of time.

This paper is organized as follows. Section 2 presents a derivation of the
dimension of the universe and its link with Lorentz invariance. Section 3
discusses the construction of a QGUT based on the $T^3$ lattice with handles.
Section 4 contains the discussion and suggestions for further investigations.

\section{The Dimension of the Universe, Lorentz Invariance and the Superposition Principle}

The invariance of the speed of light should hold in any quantum field theory.
In paperI it was shown that the loop as representative of the superposition
principle leads to $T^3$ as the fundamental Planck scale object. Since the
three-torus bounds a Lorentz four-manifold with SL(2;C) spin structure,
i.e.~is {\it nuclear}, the superposition principle {\it implies} Lorentz
invariance.

Rather than impose the dimension of space-time one would want to derive it from
first principles. In a loop-topological sense, field interactions correspond to
loops which are linked. Clearly one can exclude spatial dimensions $\le 2$ on
these grounds.
Furthermore, any two linked loops embedded in a space of dimension $\ge 4$ are
homeomorphic to two separate loops. Given that Lorentz invariance implies time
dilation of the arbitrary intersection point of two loops, it follows that
{\it the
only spatial dimension of the universe consistent with quantum mechanics is 3}.
As a corollary one finds that the occurrence of spin $1/2$ in the universe,
a property of the three-torus, follows from the superposition principle.
Since both Lorentz invariance and the dimension of the universe derive from the
superposition principle, the fact that life depends on quantum-mechanical
processes in an essential way provides a natural formulation of the Anthropic
Principle. Conversely, {\it the sum of the prime quanta}, $1+3+7=11$, {\it
implies that the group theoretic dimension of the universe is eleven}. This
dimension pertains to those scales and energies for which the GUT is realized
and direct interactions between three-tori and handles through matter degrees
of freedom can be neglected (paperI).

\section{Quantum Gravity, Superstrings and Topological Dynamics}

\subsection{K3 as a Homotopically Trivial Internal Space}

The prime quantum of $T^3$ equals 7 which implies that 4 dimensional degrees
of freedom exist on the Planck scale which are dynamically trivial under the
action of the loop algebra, generated by $T^\dagger$ and $T$, which acts on the
three-torus. This leads to the decomposition
$$G^7=T^3\times M, \eqno(1)$$
where $M$ has dimension four and $G^7$ is a manifold with a group theoretic
dimension of seven. Furthermore, $M$ is not necessarily
confined to the Planck scale, just like the homotopically trivial
three-sphere $S^3$ and the handle manifold $S^1\times S^2$. Because there
is only one type of loop and $T^{\dagger}T^3=T^4$ is the only four-manifold
which can be constructed from the three-torus under the loop algebra (paperI),
$M$ must be of the form $S(\oplus_iT^4_i)$.
Here $S$ denotes a ``simplification'' which renders M
simply-connected. The operator $S$ is local and should be a projection,
$S^2=S$. This amounts to identifying all points $x$ and $-x$ generated
by the modular
group SL(4;Z) of $T^4$, where the $x$ are coordinates local to each four-torus.

The manifold thus constructed is the compact, simply-connected Kummer K3
quartic surface in $CP^3$ with an Euler number of 24. This space is
known to be K\"ahler with vanishing first Chern class and to admit an
(anti-)self dual metric. In fact, K3 has no symmetries at all. If one views
interacting superstrings as being built up from the surfaces of three-tori,
then heterotic string duality in seven dimensions results
from the fact that
the effective compactification space changes from K3 in M-theory to $T^3$.
Also, the internal space K3$\times T^3$ has an SU(2) holonomy, and
supports $N=4$, Osp(4/4), supersymmetries with
192+192 massless degrees of freedom, i.e.~compactified 11-dimensional super
gravity[5]. The natural scale for K3$\times T^3$ is therefore the GUT scale.

\subsection{QGUT Phenomenological Preliminaries}

In paperI it was shown that the three-torus is the fundamental Planck scale
object. It was also found that the {\it non-irreducible} handle manifold
(a black hole) can exist on
all length scales and that interactions between $T^3$ and $S^1\times S^2$ occur
as matter traverses the wormholes. The three-torus has 3 branches and it is
easy to see that individual branches yield spin 1 particles and pairs of
branches lead to spin 1/2[4]. As a handle attaches itself to two singlets or
connects two branches of
a doublet it effectively converts 1-bosons into 1/2-fermions and
vice versa. The same effect occurs through the junction points of two
three-tori connected through three-ball surgery.
The three branches then yield three pairs of leptons and three doublets of
quarks. It follows that the
maximum non-integer spin carried by the three-torus, and therefore by
K3$\times T^3$, is 3/2 (i.e.~a gravitino).

In fact, In paperI it was shown that the lattice junctions support O(n) and
SU(n) symmetry groups. From the fact that the effective, group theoretic
dimension of the GUT is 11,
it follows that the dimension of these junction groups is 10, i.e.~SU(5) and
O(10). That is, one dimension is lost through projection onto an $S^1$ space
obtained through connecting junction points on the $T^3$ lattice along closed
paths. These local closed paths are needed to define metric connections and
covariant derivatives.

For supersymmetry in the presence of quantum-gravitational effects,
i.e.~$Q$-supersymmetry, the $S^1\times S^2$
number density on the $T^3$ lattice should be close
to unity. It was already shown in paperI that this density depends on the
formation, merging and evaporation of black holes as well as the rate of
expansion of the universe. From the discussion on the cosmological constant in
the Introduction it then follows that {\it the
$T^3$ field theory without handles yields $\Lambda =0$, as found in
conventional superstring theory}.
In summary, {\it the fundamental structure for the QGUT is a $T^3$ lattice
with handles}. This structure will be referred to as being {\it charged}
by $S^1\times S^2$.

\subsection{QGUT Construction}

\subsubsection{The Fundamental Topological Manifold}

Only odd sums of nuclear primes bound Lorentz manifolds (paperI).
Note that every orientable manifold in two and three dimensions is a spin
manifold and that one requires equal numbers of bosonic and fermionic fields.
Supersymmetry also requires the action to be symmetric in fermion-boson and
boson-fermion interchange processes. This requires three handle manifolds per
pair of three-tori. That is, two branches of each $T^3$ are connected
internally and the third handle links the remaining two branches of the
three-tori. So {\it at the QGUT scale a supersymmetric
manifold can be written as}
$$Q=2T^3\oplus 3S^1\times S^2,\eqno(2)$$
which assures nuclearity and Lorentz invariance.

If one works on the conformally invariant boundary $\partial Q=TQ$ of
this structure, then $\partial T^3=T_3$
is identified with a cubically interacting, gauge string field $A$.
The loop annihilation operator reduces the handles to $S^2$ spheres, which are
trivial in the connected sum. The string connection $\omega$
lives on the junction of $T_3\oplus T_3$. This yields a covariant
derivative $D=d+\omega f$ with $f$ the structure constants.
The effective geometric string action, the Chern-Simons form,
immediately follows, $S=A\times DA + 2/3A\times A\times A$, where the Verma
indices and the anti-symmetric three-tensor $\epsilon$ have been suppressed.
This suggests that M-theory on $S^1$ is dual to the
10-dimensional type IIa superstring.

\subsubsection{The Equation of Motion for Constant Charge}

In paperI an elaborate derivation was given for the topological dynamics of
space-time in terms of the $T^3$, $S^1\times S^2$, and $S^3$ prime manifolds
when the loop is the fundamental object.
The derivation in paperI was based on 1) the existence of homotopic classes of
paths in a topological manifold as
representatives of the linear superposition principle, and 2) the validity
of CPT represented by loop creation and annihilation processes.
{\it The homotopy and geometry of paths through the $T^3$
lattice, modified by the handle interactions, are here seen as different
realizations of the fundamental topological object $Q$}.
As such, the derivation of paperI is equally applicable to the loop properties
of the topological four-manifold $Q\times R$.
The QGUT equation of motion then follows from the
single time topological dynamics equation derived in paperI for a fixed
$Q\times R$ topology described by the paths in a set of connected prime
manifolds ${\cal T}_j$,
$$\sum_j [T{\cal T}_i,T^{\dagger}{\cal T}_j]/{\cal T}_j=
R(TT^{\dagger}+T^{\dagger}T){\cal T}_i,\eqno(3)$$
with the identifications
$(T,T^{\dagger}\rightarrow\partial_\mu ,\partial^\mu)$,
${\cal T}_i\rightarrow q_\lambda$. Because the topology is fixed now, the
dynamics of the theory, the left hand side in Equation (3), must be a
{\it pure self-interaction} with no reference to any topological basis set
(see paperI). This yields for the four vector $q_\lambda$
$$q^\mu [q_\lambda\Box q_\mu -q_\mu\Box q_\lambda ]=
2R\Box q_\lambda,\eqno(4)$$
with $\Box$ the four-dimensional Laplace operator or d'Alembertian. Here $R$
is a constant associated with the dynamical changes in the topology of the
charged $T^3$ lattice (see paperI). It follows immediately through
multiplication with $q^\lambda$ that a QGUT is only self-consistent if the
charged $T^3$ lattice has $R=0$. One finds that {\it the resulting
self-interaction is cubic as, demanded by the triple loop structure on the
three-torus, and linear in the second derivative operators}.
A natural interpretation for the quadruplet is that
it describes the excited states of Q, analogous to the vibration modes of the
superstring. The possible functional forms of $q_\lambda$ are determined by
the quadruplet solutions to the equation of motion, {\it and the cyclic
structure of} $Q$.

The tensor $q$ of rank one results as follows. The three {\it loop
homotopic} degrees of freedom of the seven-dimensional internal space
K3$\times T^3$ are equivalent to a tensor $V$ of rank 3 and dimension 7.
This tensor should be symmetric in two of its indices because it describes
three-point interactions and two of the $T^3$ legs
are connected internally through $S^1\times S^2$ handles. So $V$ supports
196 algebraically independent components.
From the 196 degrees of freedom, 192 (on each three-torus) can be
associated directly with
11-dimensional, N=1 super gravity compactified on K3$\times T^3$, i.e.~the
low energy limit Osp(1/4). The remaining four degrees of freedom
are purely quantum-gravitational in nature and correspond to the four currents
(wave amplitudes) which can flow along the homotopically inequivalent
paths associated with the charged $T^3$ lattice.

The equation of motion for the quadruplet $q_\lambda$ on $Q\times R$ is now
$$q_\lambda q^\mu \Box q_\mu =q^\mu q_\mu\Box q_\lambda,\eqno(5)$$
{\it in the constant charge limit} (see \S 3.3.3 below).
A large class of solutions of (5) is determined by the wave
equations $\Box q_\lambda =0$. This equation of motion is of the form
``the boundary of the boundary is zero'', in
analogy with the sourceless Maxwell equations. Its solutions are represented
by the well-known traveling wave forms. Another class of solutions is
determined by the Klein-Gordon equations $\Box q_\lambda +m^2q_\lambda =0$.
Because these solutions are applicable to a constant charge system, they
also hold for the neutral $T^3$ lattice.

Note that Equation (5) is invariant under CPT, a property
which is conserved in the full equation of motion derived below.
In PaperI it was shown that the SL(2;C) gauge group follows
naturally from a nuclear manifold. The theory is therefore manifestly
Lorentz invariant with $\partial^\mu =\eta^{\mu\nu}\partial_\nu$.
Furthermore, the Dirac matrices can be constructed from
the hermitian Pauli matrices contained in SL(2;C), and yield a local
formulation of spin 1/2.

\subsubsection{Symmetry Breaking and the Equation of Motion in the Presence of Charge Fluctuations}

The properties of the resulting QGUT depend on both the charge on
the $T^3$ lattice as well as the energy density of the
matter degrees of freedom. The interactions in the GUT on $T^3$, in the low
charge and low energy limit, correspond to
11-dimensional N=1 super gravity compactified
on K3$\times T^3$. That is,
Osp(N/4)$\supset$O(N)$\times$Sp(4) with N=4
for the massless sector in dimension 4.
For small charge and large energy (above the GUT scale), M-theory results
on K3. That is, M-theory corresponds to the small charge quadruplet solutions
on the $T^3$ lattice. Membranes appear to have the correct limiting behavior
for small energy and encompass all the
known types of superstrings. For decreasing charge and energy density,
the QGUT domains will therefore become interspersed with regions described
by M-theory and 11-dimensional super gravity.

For large charge densities, $Q_{\rm h}\approx 1$, but with quantum
perturbations in the local number of handles, which are subject to evaporation,
an additional field is
generated. These fluctuations in the charge on the $T^3$ lattice cause phase
changes in the currents flowing through $Q$.
{\it The fundamental object to solve for is therefore}
${\rm e}^{i\phi}q_\lambda$, with $\phi$ a function of time and
position. The presence of this
U(1) multiplier then leads to the {\it full QGUT equation of motion}
$$2i\partial_\nu\phi [(\partial^\nu q_\lambda )q^\mu q_\mu -(\partial^\nu
q_\mu )q^\mu q_\lambda ]=q_\lambda q^\mu\Box q_\mu -q^\mu q_\mu \Box
q_\lambda ,\eqno(6)$$
with the scalar constraint $q^\mu q_\mu =\phi^2$ to preserve locally the
internal phase information of the system.
PaperI discusses the topological formulation of this constraint in terms of
multiplication and the effective interaction potential.
This conservation law remains in effect as the QGUT on $Q\times R$ is broken
down to an M-theory. The internal K3$\times T^3$ space only becomes accessible
for $Q_{\rm h}<<1$ and small energy. Recall here that K3 has no symmetries at
all, and therefore K3$\times T^3$ supports non-trivial holonomies only when
$Q$-supersymmetry is replaced by the gauge group Osp(N/4).
Because the evolution of $\phi$ is driven by the handle degrees of freedom,
it follows that {\it in M-theory the field $\phi$ obeys the condition
$\partial_\nu\phi =0$, and is effectively frozen in}. Note that the handle
quantum fluctuations lead to a fourth-order interaction term which couples the
quadruplet and the scalar field.
Note that there are precisely enough
degrees of freedom and constraints in the theory to support both the
quadruplet and a scalar field. In fact, multipliers other than U(1) would
yield the system of equations underdetermined. The {\it full} solution is now
determined by Equation (6), the scalar constraint, and the cyclic properties
of $Q$.

The junction potential has the form
$V(q^\mu q_\mu )=\lambda(q^\mu q_\mu )^2+\mu^2q^\mu q_\mu$.
This potential follows from the
four-fold homotopic symmetry associated with the $T^3$ lattice and reduces to
the form of paperI under the scalar constraint.
In the low charge limit, the field $\phi$ has a constant, Lorentz invariant
vacuum expectation value $<0|\phi (x)|0>=c\not =0$, defined on the junctions
of the $T^3$ lattice.
{\it The additional Poincar\'e scalar is, in fact, the Higgs field}.
The vacuum expectation value of the Higgs field is non-zero for $\mu^2<0$
which leads to
spontaneous symmetry breaking, as first suggested by Nambu and co-workers.

\subsubsection{Topological Defects}

As the charge on the $T^3$ lattice decreases it need not do so in a homogeneous
manner. Therefore, topological defects can result where the phase $\phi$
differs between charged and neutral $T^3$ lattice sites.
These topological defects have no electrical charge and spin 0.
It is proposed here that {\it these topological defects contribute to
the so-called dark matter necessary in models of large scale structure
formation}.

The number of degrees of freedom of $Q$ is given by
$$N_Q=(TT^\dagger +T^\dagger T)(2T^3\oplus 3S^1\times S^2)=23.\eqno(6)$$
In the loop homotopic approach adopted here, these degrees of freedom are
all equivalent and equipartition should apply. For the neutral lattice manifold
$P=T^3\oplus T^3$, one has $N_P=14$. The latent heat is therefore
$$H=(N_Q-N_P)m_{\rm Planck}/N_Q=9m_{\rm Planck}/23.\eqno(7)$$
Thus, one finds
$$m_q\le H/2=9m_{\rm Planck}/46\eqno(8)$$
for the mass of a topological defect $q$ produced by the phase difference
across the junction of $P$. The determination of the precise mass distribution
of the defects, which is important for the small scale power spectrum,
requires a more detailed computation and will be presented
elsewhere. The efficiency of this defect mechanism depends on the
number of (mini) black holes in the universe at early epochs as discussed in
paperI. Furthermore, the characteristic energy scale for a supersymmetric GUT
is of the order of $m_{\rm GUT}=m_{\rm Planck}/N_Q$, which is smaller than the
typical mass of a defect.
It follows that approximately $H/m_{\rm Planck}\approx 40$\% of the dark
matter in the universe can be in the form
of scalar topological defects. Additional GUT phase transitions can increase
the percentage of non-baryonic dark matter. Still, it appears that current
estimates of the fraction of baryonic matter in the universe [6] suggest that
not all dark matter in the universe is non-baryonic if $\Omega =1$.

\subsubsection{Quantum Gravity Effects and Large Scale Structure}

From the above discussion it follows that the QGUT energy scale corresponds to
a size and matter density of the universe where mini black holes are
formed rapidly enough to sustain the topological manifold $Q$.
During the QGUT epoch the formation and evaporation of mini black holes
leads to spatial variations in the scalar field $\phi$ which are frozen in
when the charge on the $T^3$ lattice becomes small.
The multiplication effect of paperI then assures that the associated
(thermal) fluctuations in the quantum fields on the charge-free $T^3$ lattice
are suppressed and scale free.

The strength of these initial, large scale perturbations can be
computed rigorously with very few assumptions.
The quadruplet interaction is intrinsically cubic. Therefore, the 23
equipartitioned degrees of freedom on $Q$ lead to a maximum amplitude which
is given by
$$\delta T/T=N_Q^{-3}=8.2\times 10^{-5}.\eqno(9)$$
This amplitude is the first non-zero correction term to the average energy
of $Q$ on purely thermodynamic grounds.
If the $1\sigma$ topological correction due to multiplication
(paperI) is applied one finds a final amplitude $\delta T/T=3.7\times 10^{-5}$
on the horizon scale.
This amplitude is consistent with Cosmic Microwave Background
measurements if the soft equation of state during recombination at
$z\sim 1000$ is included[7].

\section{Discussion}

It has been shown that the requirement of Lorentz invariance follows from the
superposition principle and leads to a spatial dimension of the universe equal
to three. An equation of motion has been derived for a QGUT on the
supersymmetric manifold $Q=2T^3\oplus 3S^1\times S^2$, i.e.~a
charged $T^3$ lattice.
M-theory results for the low charge limit. For small energy densities, the
theory is described by 11-dimensional super gravity and the compactification
procedure is prescribed by the QGUT.
It has been found that the non-baryonic mass content of the universe
is directly linked with Planck scale dynamics through the formation and
evaporation of mini black holes and the production
of scalar topological defects of mass $m_q\le 9m_{\rm Planck}/46$ during the
QGUT era.

The general supersymmetric solution for the QGUT and M-theory, including the
cosmological evolution of the defects and the choice of initial conditions for
the equation of motion on an evolving $T^3$ lattice (paperI), will be
presented in [8]. That is, the state space will be constructed from the
quadruplet solutions and the cyclic properties of $Q$.
The nature of black hole singularities can then be addressed as well, since
they are quadruplet solutions in the low charge limit. In particular, such
solutions should provide a statistical mechanical formulation of black hole
entropy.

The link between black holes and Planck scale phenomena was already suggested
in paperI, and it has been investigated here how very small and very large
scale physics are linked. The existence of a charge
threshold at which the QGUT symmetry is broken, is equivalent to a quiescent
and stable quantum foam which derives from a relatively small set of
macroscopic black holes. {\it This macroscopic set of black holes, as a
measure of the topology of the universe, fixes the Planck scale quantum
fluctuations, and the nature of field interactions}.

The author is indebted to G.~van Naeltwijck van Diosne, J.A.A.~Berendse-Vogels
and W.G.~Berendse for valuable inspiration, and to C.~M.~Carollo and
F.~Kemper for stimulating discussions.
The author acknowledges with gratitude the support of NASA grant
NAGW-3147 from the Long Term Space Astrophysics Research Program.


\begin{thebibliography}{}

\bibitem{}
S. Kaku, Introduction to Superstrings (Springer-Verlag, 1990).

\bibitem{}
S.W. Hawking and A. Strominger 1991 in Quantum Cosmology and Baby
  Universes, eds. S. Coleman, J.B. Hartle, T. Piran \&\ S. Weinberg,
  (World Scientific, 1991) p.\ 245, p.\ 272.

\bibitem{}
M. Spaans, Nuc. Phys. B in press (1997, paperI).

\bibitem{}
J.A. Wheeler in Quantum Cosmology, eds. L.Z. Fang \&\
  R. Ruffini, (World Scientific, 1987) p.\ 27.

\bibitem{}
M.J. Duff, B.E.W. Nilsson and C.N. Pope, Phys. Lett. B 129 (1983) 39.

\bibitem{}
P.J.E. Peebles, Principles of Physical Cosmology (Princeton:PUP, 1993).

\bibitem{}
C.A. Norman and M. Spaans, ApJ submitted.

\bibitem{}
M. Spaans, in preparation.

\end{thebibliography}
\end{document}